\documentclass[aps,pre,twocolumn,showpacs,floatfix,preprintnumbers,amsmath,amssymb]{revtex4}
\usepackage{epsfig}
\usepackage{graphicx}
\usepackage{dcolumn}
\usepackage{bm}

\begin{document}
\title{Zero delay synchronization of chaos in coupled map lattices}
\author{M. S. Santhanam and Siddharth Arora}
\affiliation{Physical Research Laboratory, Navrangpura, Ahmedabad 380 009, India.}

\begin{abstract}
We show that two coupled map lattices that are mutually coupled to one
another with a delay can display zero delay synchronization if they
are driven by a third coupled map lattice. We analytically estimate the
parametric regimes that lead to synchronization and show that the presence of
mutual delays enhances synchronization to some extent.
The zero delay or isochronal synchronization is reasonably robust against
mismatches in the internal parameters of the coupled map lattices and
we analytically estimate the synchronization error bounds.
\end{abstract}
\maketitle
\section{Introduction}
Synchronization is one possible form of emergent dynamics
displayed by coupled oscillator systems and this is seen
in a wide variety of physical phenomena, e.g, synchrony among neural activity
in the brain \cite{glass}, intensity of coupled lasers to light pulses emitted
by fireflies \cite{kurth1}.
It is now well established that two chaotic systems,
with appropriate coupling, can exhibit synchronized behavior \cite{pc,kurth2}.
Synchronized chaotic dynamics has been reported in systems consisting
of well known models
of chaos like the logistic map, Lorentz system, Henon map, coupled map lattices etc.,
which represent a wide collection of discrete and continuous time systems.
The growing interest in chaos and synchronization is
partly due to its potential applications in chaos control, chaos based cryptography,
neural networks and biological systems.

Much of the work on synchronization has concentrated on instantaneous
coupling of the dynamical systems \cite{kurth1}. This implies that we disregard the
finite time it takes for the interaction or the information to travel
from one system to the other. Consider two identical chaotic systems
represented by $x_{n+1}=F(x_n)$ and $y_{n+1}=F(y_n)$, started from different
initial conditions. To synchronize their solutions, $x_n$ and $y_n$, we suitably
couple both the systems. Then the modified equations will be
$x_{n+1}=F(x_n) + g_1(y_n)$ and $y_{n+1}=F(y_n) + g_2(x_n)$. In this form of
coupling,
we have implicitly assumed that $y_n$ is instantly available to the
$x$-system without any delay and vice versa. This cannot be true in general.
Many physical phenomena that display synchronization are often spatially separated
and the time taken for the information to travel is not negligible.
For example, synchronization of neuronal activity in the brain involves
time delays due to information processing and transfer between different
parts of the brain \cite{bdelay} and is estimated to be about tens of
milliseconds \cite{bdelay1}. In the context of using chaos synchronization for
secure communication, it is usual for the sender and
receiver to be spatially separated and information takes finite time to
travel between them. Then, the question is, can the delayed interactions
lead to synchronization of coupled systems ? The works done in the last
few years show that delayed couplings can lead
to synchronization \cite{masoller,atay} as well as new scenarios such as amplitude death
in limit cycle oscillators and coupled oscillators \cite{reddy}, multi stable
synchronization \cite{multis} and symmetry breaking \cite{symmb}. Techniques for
controlling pathological rhythms in neurons based on delayed feedback have also been
reported \cite{rose}.

The experiments on information processing in the brain and neurosciences 
are providing evidence for new features of delayed synchronization, namely
that of near zero delay synchronization of signals from spatially
separated regions \cite{eeg-rev}. It has been
reported that spatially separated cortical regions in the brain
of the cat display synchronization {\it without any lag} \cite{cat}.
Neuronal firings from left and right cortex regions recorded on
primates show near zero synchrony maintained over
considerable distances \cite{murthy}. The idea of spatially distributed systems
synchronizing without delay continues to attract research attention since the mechanism
leading to such an effect is not yet clear and continues to be debated \cite{free}.

Recently one mechanism for zero delay synchronization of mutually
coupled oscillators has been demonstrated experimentally in a system
of three semiconductor lasers \cite{fischer}. In this case, a central driving laser L$_2$
is bidirectionally coupled with two other mutually delay coupled lasers,
L$_1$ and L$_3$. Then, the
delay coupled lasers L$_1$ and L$_3$ display zero delay or isochronal synchronization and
is shown to be reasonably robust. A variant of this scheme has been used
to propose a method for bidirectional secure communication using delay
coupled oscillators \cite{rroy}. Simultaneously, the modeling and analysis of zero
lag synchronization maintained over large distances in neurons is beginning to
take shape \cite{lagmodel}.

In this paper, we show that zero delay synchronization
can be achieved in delay coupled spatially extended systems, namely a
coupled map lattice (CML), if they
are driven by a third such system. This application is motivated by
the fact that the activity in the cerebral cortex of the brain (as measured
by electroencephalograph, for example) is due to the interaction
between millions of neurons that are spatially distributed. The
zero lag synchronization occurs between groups of such spatially
distributed neurons. Hence one is led to
consider a collection of coupled oscillators that are spatially
separated. Another important application is in the area of 
secure communication. Coupled map lattices can be applied to
encryption of messages in multi-channel communication \cite{mcc}.
In real-time, for multichannel communications that require
security, there is a need
for as many different chaotic signals as the number of channels to encode
the messages sent in each of the channels. Signals in these channels
are encoded by the chaotic time series from one of the lattice points of the CML.
The CML, being a high dimensional
chaotic system, provides sufficient security against most attacks. Presence of
chaotic, zero delay synchronization would allow
the receiver to decode the message in all the channels at the same time. For real-time
applications synchronization must be achieved in shortest possible time
and zero delay synchronization is ideally suited for this purpose.
For a review of certain aspects of synchronization in spatially extended systems
and its applications, see ref \cite{zan}.

In the next section, we briefly review the coupled map lattice paradigm
and introduce our model. Further, in subsequent sections, we report results
on zero lag synchronization from this model, obtain analytically the parametric
regimes where this occurs and also bounds for synchronization errors due to
parameter mismatches.

\section{Coupled Map Lattice}
We consider the coupled map lattice given by,
\begin{equation}
x_{n+1}^i = (1-\epsilon) f[x_n^i] + \frac{\epsilon}{2}
  \left( f[x_n^{i-1}]  + f[x_n^{i+1}] \right)
\label{cml}
\end{equation}
where $i=1,2 ....L$ is the index for the lattice site
and $\epsilon$ is the coupling strength parameter.
This was originally introduced \cite{cml,cml-app} as a model for
chaos in spatially extended systems. There have been attempts to
model real life phenomena based on CMLs \cite{cml-app}.
They display a rich variety of dynamical
regimes ranging from frozen random patterns to spatio-temporal
chaos upon variation of the parameters.
We use periodic boundary conditions, so that $x_n^{L+1}=x_n^{1}$ leading
to a ring type lattice. 
Here, the local dynamics uses the logistic equation, $f[x]=a x(1-x)$, where
$a$ is the chaos parameter. In Sec III(A), we will denote this map
showing its explicit parameter dependence as $f[x,a]$.
\begin{figure}[h]
\includegraphics*[width=8.8cm]{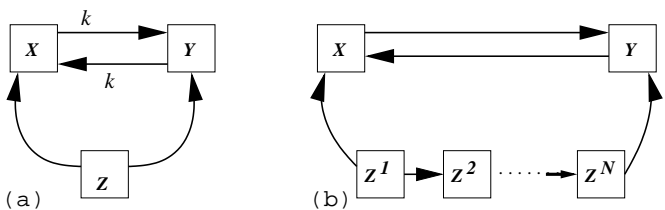}
\caption{Coupling scheme of CMLs. (a) $x$ and $y$ are mutually
delay coupled CMLs. The $z$-CML is the driver. (b) The driver is
a collection of $N$ unidirectionally coupled CMLs. $k$
denotes the delay time in mutual coupling.}
\label{scheme}
\end{figure}
In this work, we are attempting to synchronize two CMLs labeled $x$ and $y$ at
zero delay when both are mutually delay-coupled and are driven by a third CML
labeled $z$. This is schematically
shown in Fig. \ref{scheme}(a). The second CML labeled
$y$ is obtained by replacing $x$ with $y$ in Eq. (\ref{cml}); CML $z$ can
be obtained in a similar way. We mutually couple 
$x$ and $y$ CMLs with a delay.
The $z$-CML is the driver and it is unidirectionally coupled to $x$ and $y$ CMLs.
It is given by,
\begin{equation}
z_{n+1}^i = (1-\epsilon) f[z_n^i] + \frac{\epsilon}{2}
  \left( f[z_n^{i-1}]  + f[z_n^{i+1}] \right)
\label{cml-z}
\end{equation}
The modified form of $x$-CML is given by,
\begin{eqnarray}
x_{n+1}^i & = & \Gamma \left\{ f[x_n^i] + \beta f[y_{n-k}^i] + \alpha ~f[z_{n}^i] \right\} \nonumber \\
     &  & + ~~\frac{\epsilon}{2} \left\{ f[x_n^{i-1}]  + f[x_n^{i+1}] \right\}
\label{cml-x}
\end{eqnarray}
where $\Gamma = (1-\epsilon)/(1+\alpha+\beta)$. Similarly, the $y$-CML is also modified
and becomes,
\begin{eqnarray}
y_{n+1}^i & = & \Gamma \left\{ f[y_n^i] + \beta f[x_{n-k}^i] + \alpha ~f[z_{n}^i] \right\} \nonumber \\
     &  & + ~~\frac{\epsilon}{2} \left\{ f[y_n^{i-1}]  + f[y_n^{i+1}] \right\}.
\label{cml-y}
\end{eqnarray}
The Eqns. (\ref{cml-z}-\ref{cml-y}) represent our coupling scheme shown in 
Fig. \ref{scheme}(a) for zero lag synchronization.
The parameters $\beta \ge 0$ and $\alpha \ge 0$ represent the strength of delayed mutual
coupling and the strength of coupling with the driver CML respectively.
Note that in Eqns. (\ref{cml-x},\ref{cml-y}), delay $k$ is introduced in
the mutual coupling term. If $\alpha=0$ (absence of drive CML), then no isochronal
synchronization takes place between $x_n(i)$ and $y_n(i)$ in the presence of
mutual delays between them. However, if $\beta=0$
these CMLs synchronize beyond some critical value of $\alpha$. We will
explore the general case when $\alpha , \beta \ge 0$, $k>0$ and show numerical
evidence for zero lag synchronization of $x$- and $y$-CMLs but not with $z$-CML.
The coupling parameter $\epsilon$ and $a$ are chosen such that CML generates chaotic motion.

The coupling scheme shown in Fig. \ref{scheme}(a) is reminiscent of
the generalized synchronization (GS) that has been widely studied in the
last one decade \cite{aba}. In GS scenario, there is one driver and
a driven (response) system and the state of the latter depends on the
former. This is the likely case in the absence of mutual coupling, {\it i.e.}, $\beta=0$.
However, in this scheme we are considering one driver and
two driven systems which themselves are mutually coupled to one another
with a delay. It is also relevant to point out that in this scheme the driver
and the driven system need not necessarily be one unit each but can be
composed of many (sub)systems as shown in Fig. \ref{scheme}(b). In this scheme,
instead of one driver CML, a collection of $N$ unidirectionally coupled
CMLs are used to drive the $x$ and $y$ CMLs.  The detailed
results for this scheme will be presented elsewhere.

\begin{figure}
\includegraphics*[width=8.8cm]{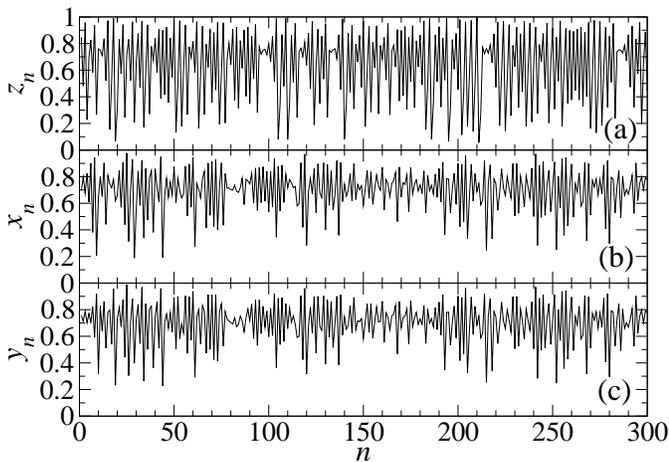}
\caption{(a) $z_n$ (b) $x_n$ and (c) $y_n$ for 525th lattice site of the CML system
in Eqns. (\ref{cml-z}-\ref{cml-y}). The local map
parameter is $a=4.0$. The local coupling strength in CMLs
is $\epsilon=0.1$ and $\alpha=\beta=1.0$. The $x$ and $y$ CMLs are
mutually coupled with delay $k=26$.}
\label{fig2}
\end{figure}
\begin{figure}
\includegraphics*[width=8.8cm]{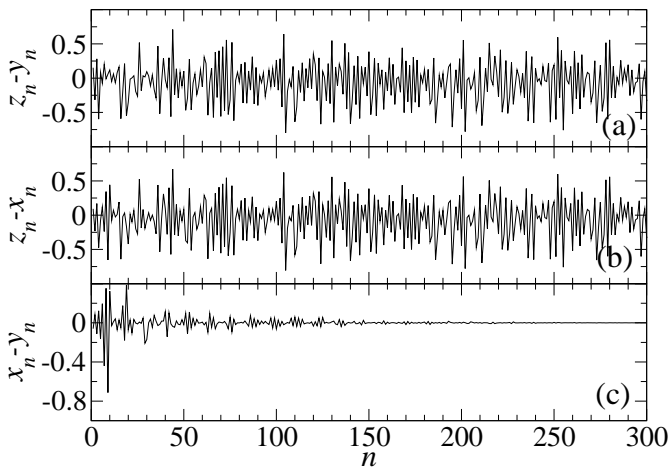}
\caption{(a) $z_n-y_n$, (b) $z_n-x_n$ and (c) $x_n-y_n$. The parameters
are the same as in Fig. \ref{fig2}.}
\label{fig3}
\end{figure}
\begin{figure}
\includegraphics*[width=8.8cm]{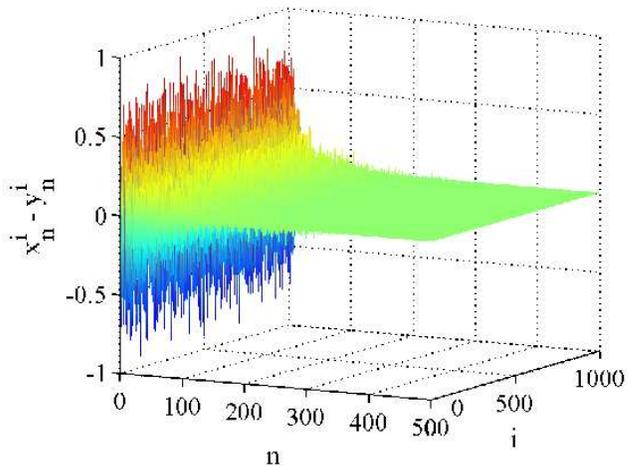}
\caption{(Color Online) The quantity $x_n^i-y_n^i$ plotted as function
of $i$ and $n$. Notice that the region beyond $n > 100$ is
flat indicating synchronization without delay in the entire coupled
map lattice. The parameters are the same as in Fig. \ref{fig2}.}
\label{fig4}
\end{figure}

\section{Zero delay synchronization}
The coupled map lattices in Eqns. (\ref{cml-z}-\ref{cml-y}) with
$L=1000$ lattice elements are iterated for 6000
discrete time steps. Each of the CML is initialized at $n=0$ with a
different realization of uniformly distributed random numbers.
The logistic map parameter is $a=4.0$ such that the local map
dynamics is chaotic and the local coupling strength is $\epsilon=0.1$. For this combination
of $a$ and $\epsilon$, the CML in Eq. (\ref{cml}) is known to display
spatio-temporal chaos \cite{cml-app}. The $x$ and $y$ CMLs are mutually coupled
with a delay of $k=26$. The $z$ CML drives both the $x$ and $y$ CMLs.
In Fig. \ref{fig2}, we show a typical
time series for $x_n, y_n$ and $z_n$ drawn from 525th lattice point
from the system of CMLs given by Eqns. (\ref{cml-z}-\ref{cml-y}).
Notice that each of them is chaotic and beyond 100th time step
$x_n$ and $y_n$ are synchronized without delay. The difference
between the pairs of time series is shown in Fig. \ref{fig3} and
clearly the delay-coupled CMLs, $x$ and $y$, exhibit isochronal synchronization
(Fig. \ref{fig3}(c)) but they do not synchronize with the driver
$z$-CML as seen in Fig. \ref{fig3}(a,b). Even though results from a typical time
series from one lattice point
is displayed in Figs. \ref{fig2} and \ref{fig3}, we observe
synchronization for all the lattice points of $x$ and $y$ coupled map
lattice. This is shown in Fig. \ref{fig4} as a space-time plot which has
a flat region coinciding with zero of $z$-axis for $n>100$ at every lattice point.

\begin{figure}
\includegraphics*[width=8cm]{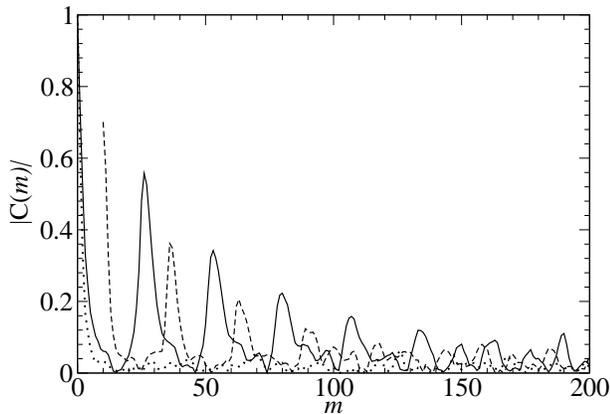}
\caption{Absolute cross-correlation $|C(m)|$ as a function of lag $m$.
$|C(m)|$ between $x_n$ and $y_n$ (solid line) and between $x_n$ and $z_n$
(dashed line) taken from their 525th lattice point is displayed. The dashed
line is shifted by 10 units along $x$-axis for clarity. The 
cross-correlation between $y_n$ and $z_n$ is almost indistinguishable
from the dashed line since $x_n$ and $y_n$ are in synchrony. If mutual
coupling is absent ($k=0$), then recurrences are not seen (dotted line).}
\label{fig5}
\end{figure}

The isochronal synchronization of CMLs demonstrated in Fig. \ref{fig4}
depends on the strength parameters $\alpha$ and $\beta$. Results presented here
indicate that there is a critical value of $\alpha$ and $\beta$,
other parameters remaining the same, below which synchronization
does not take place. However, the time taken to achieve
synchrony is found to increase with magnitude of
the delay $k$. The effect of various parameters on synchronization is
the discussed in the next section.
In order to understand the correlations that exist between
$x_n$, $y_n$ and $z_n$ at $i$th lattice site, we study the
lagged cross-correlation defined as,
\begin{equation}
C(m) = \frac{\sum_{n=1} \left( x_n - \overline{x} \right) 
                  \left( y_{n+m} - \overline{y} \right)}{\sigma_x \sigma_y}
\label{crosscor}
\end{equation}
where $\bar{x}$ and $\bar{y}$ are the sample means, $\sigma$ the corresponding
standard deviation and $m$ represents the lag
and lattice site index is suppressed. In Fig. \ref{fig5}, the solid line shows the
lagged cross correlation between the iterates of
$x$ and $y$ CMLs at 525th lattice site. At zero lag, $x_n$ and $y_n$ are
almost perfectly correlated with $|C(0)|=0.998$ indicating perfect
synchronization without delay. Along with this, the mutual coupling between
$x$ and $y$ CMLs
with delay $k=26$ leads to partial recurrences in $|C(m)|$ at similar intervals.
In contrast, for $x_n$ and $z_n$ (dashed line in Fig. \ref{fig5}), $|C(0)|=0.701$
indicates the absence of identical synchronization (see also Figs \ref{fig3}(a,b)).
Due to synchrony between $x_n$ and $y_n$, again similar
result holds good for $y_n$ and $z_n$, with peaks in $|C(m)|$ separated by delay $k=26$.
If we apply instantaneous coupling, i.e, $k=0$ then the recurrences would
be absent, as shown by the dotted line in Fig. \ref{fig5}, indicating that $x$ and $y$ CMLs
would not maintain any memory of the dynamics of the other CML in them. Similar results
hold good for all the lattice sites in CML.

The recurrence pattern of $|C(m)|$ shown in Fig. \ref{fig5} can be used to
detect mutually delayed couplings and estimate the magnitude of delay
in real physical systems. As shown in that figure, the absence of
delayed couplings would not lead to any recurrence. It is essential that
the physical system should be composed of many subsystems which can
display synchronization among themselves. The time interval between
the peaks in $|C(m)|$ would give an estimate of the magnitude of delay.
However, in a complex physical system, the cross correlations alone will
not be sufficient in identifying the driver system. At this point, we also remark
about the possibility of replacing $z$-CML by a noise process. Do we expect
synchronization then ? In some restricted range of parameters, the same noise
process driving both $x$- and $y$-CML can lead to isochronal synchronization.
However, the neat recurrence structures and the associated memory features shown in
Fig. \ref{fig5} will be absent in such a case and the magnitude of delays will
not carry any significance.

\section{Sensitivity to parameters}
 In this section, we discuss the parametric regimes in which the synchronization
occurs. As pointed out before, the isochronal synchronization depends on the
parameters $k$, $\alpha$ and $\beta$. To understand the role of these
parameters, we apply linear stability analysis.
The synchronized solution of interest is,
\begin{equation}
u^i_{n+1} = x_{n+1}^i - y_{n+1}^i = \bar{u}_{n+1} \;\;\;\;\;\; \mbox{for all}\;\; n.
\label{syncsol}
\end{equation}
In our case, $ \bar{u}_{n+1}=0$.
We will perform a linear stability analysis about this solution
to determine the parameters that will lead to synchronization. 
For convenience, we will shift to new variables defined as,
\begin{equation}
u^i_{n} = x_n^i - y_{n}^i, \;\;\;\;\;\;\;\;\;\;\;\; v^i_{n} = x_n^i + y_{n}^i.
\end{equation}
The dynamics of $u^i_{n+1}$ can be written as,
\begin{eqnarray}
u^i_{n+1} & = & \frac{g(u_n^i,v_n^i)}{1+\alpha+\beta} - 
                 \frac{\beta (1-\epsilon)}{1+\alpha+\beta} ~g(u_{n-k}^i,v_{n-k}^i)  \nonumber \\
          & + &  \epsilon \left\{ \sum_{j \thicksim i} g(u_n^j,v_n^j) -
              \frac{g(u_n^i,v_n^i)}{1+\alpha+\beta}  \right\}.
\label{pert}
\end{eqnarray}
where $g(u_n^i,v_n^i) = f(x_n^i) - f(y_n^i) = a u_n^i (1 - v_n^i)$ and
$j \thicksim i$ represents summation over nearest neighbours.
We will follow the elegant technique discussed in Ref. \cite{joy}. The connection
topology, here being the nearest neighbour, is encoded in the spectra of
the graph Laplacian defined as, $(\Delta w)_i = (1/n_i) \sum_{j \thicksim i}
( w_j - w_i )$. We will use the fact that the eigenmodes $\phi$ of Laplacian
are obtained from the eigenvalue equation $\Delta \phi_m = -\lambda_m \phi_m$ \cite{joy},
where $-\lambda_m$ is the eigenvalue. Then, we will consider perturbations
to synchronized solution $\bar{u}_n$ in Eq. (\ref{syncsol}) by the
$m$th eigenmode $\phi^i_m$  as,
\begin{equation}
u^i_n = \bar{u}_n + \mu ~\delta u_n^m ~ \phi^i_m,
\end{equation}
such that for $\mu << 1$, $\delta u_m(n) \to 0$ as $n \to \infty$ for
synchronized solutions.
Substituting this in $g(u_n^i,v_n^i)$ and Taylor expanding it about $\mu=0$,
we get,
\begin{equation}
g(u_n^i,v_n^i) = g(\bar{u}_n,v_n^i) + \mu ~\delta u_n^m ~\phi^i_m ~g'(\bar{u}_n),
\end{equation}
and $g'(\bar{u}_n)$ should be taken to mean $g(u_n^i,v_n^i)$ evaluated at the
synchronized solution $\bar{u}_n$.
We substitute this in Eq. (\ref{pert}) to obtain,
\begin{eqnarray}
\delta u^m_{n+1} ~ \phi^i_m  & = &
\left\{ \frac{\epsilon}{1+\alpha+\beta} \Delta \phi^i_m + 
        \frac{1}{1+\alpha+\beta} \phi^i_m \right\} \nonumber \\
         & & \delta u^m_n ~ g'(\bar{u}_n) -
           \frac{\beta(1-\epsilon)}{1+\alpha+\beta} \nonumber \\
           & & \delta u^m_{n-k} ~\phi^i_m ~ g'(\bar{u}_{n-k}).
\end{eqnarray}
Using $\Delta \phi_m = -\lambda_m \phi_m$ and after some simple manipulations, we get,
\begin{eqnarray}
\delta u^m_{n+1} = \left( \frac{1-\lambda_m \epsilon}{1+\alpha+\beta} \right) 
                  \delta u^m_n ~ g'(\bar{u}_n) - \nonumber \\
  \frac{\beta(1-\epsilon)}{1+\alpha+\beta} ~ \delta u^m_{n-k} ~ g'(\bar{u}_{n-k})
\label{deleqn}
\end{eqnarray}
This is the relation we need to analyse the stability of synchronized solutions.
The eigenvalue $-\lambda_m$ is dependent only on the connection topology
of the CML. For nearest neighbour coupling, the non-zero eigenvalues of $\Delta$ are
$\lambda_m = 1 - \cos(2 \pi m/L)$, $m=1,2,......L-1$ \cite{joy}. If $L>>1$,
the largest eigenvalue is 2 if $L$ is even and $1+\cos(\pi/L) \sim 2$ if $L$ is odd.
Putting $\lambda_m=2$, Eq. (\ref{deleqn}) can be analysed for the following two cases.

\subsubsection{Absence of mutual coupling, $\beta=0$}
\begin{figure}[t]
\includegraphics*[width=5cm]{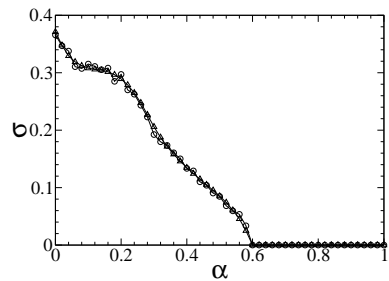}
\caption{Degree of synchronization $\sigma$ as a function of $\alpha$ for $\beta=0$
and $\epsilon=0.1$. The delays are $k=4$ (circles) and $k=24$ (triangles).}
\label{b0}
\end{figure}

Firstly, we consider the case $\beta=0$, {\it i.e}, absence of mutual coupling.
In this case, we have from Eq. (\ref{deleqn}),
\begin{equation}
\delta u^m_{n+1} = \left( \frac{1-2\epsilon}{1+\alpha} \right)
                  \delta u^m_n ~ g'(\bar{u}_n)
\label{deleqn1}
\end{equation}
The condition for local stability is,
\begin{equation}
q=\lim_{N\to\infty} \frac{1}{N}\log \frac{|\delta u^m_{N+1}|}{|\delta u^m_0|} < 0.
\label{locstab}
\end{equation}
Since $g'(\bar{u}_n) = f'(x_n)$, we can
iterate Eq. (\ref{deleqn1}) $N$ times to obtain
\begin{equation}
q = \log \left| \frac{1-2\epsilon}{1+\alpha} \right|
             + \langle \log |f'(x_n)| \rangle,
\end{equation}
where $\langle . \rangle$ denotes the time average. If the local map
$f(x_n)$ is ergodic, this average can be replaced by an
ensemble average and we have $\langle \log |f'(x_n)| \rangle = \log 2$,
the Lyapunov exponent of the logistic map. Hence, the condition for synchronization
for $\beta = 0$ turns out to be $|2(1-2\epsilon)/(1+\alpha)| < 1$, which implies
\begin{equation}
\alpha > |2 - 4 \epsilon|-1.
\label{condition1}
\end{equation}
This holds good for any value of $k$ as is to be expected.
Secondly, note that if $\alpha=0$, the stability condition in 
Eq. (\ref{condition1}) will not be satisfied for any value of $\epsilon$
for which synchronization takes place.
Hence, for $\alpha=0$, {\it i.e}, in the absence of drive CML, 
no isochronal synchronization can take place.
This condition is verified by numerical simulations of CMLs in
Eqns. (\ref{cml-z}-\ref{cml-y})
with $\epsilon=0.1$ and $\beta=0$. We define the degree of synchronization
to be, 
\begin{equation}
\sigma = \langle (x_n^i - y_n^i)^2 \rangle,
\end{equation}
where the average $\langle . \rangle$ 
is taken over all the lattice points for 50000 iterations after discarding
the initial 10000 time steps \cite{atay}. If the system synchronizes, then $\sigma \to 0$
as $n \to \infty$. The results in Fig \ref{b0} show that
for $\alpha > 0.6$, we obtain $\sigma < 10^{-20}$ leading to synchronization
and this confirms
the validity of analytical condition in Eq. (\ref{condition1}).

\begin{figure}[t]
\includegraphics*[width=8cm]{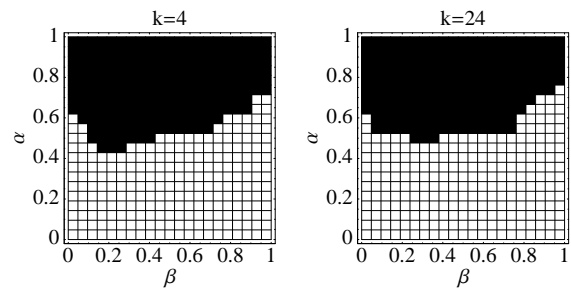}
\caption{Degree of synchronization $\sigma$, as a function of $\alpha$ and $\beta$,
obtained by numerically iterating Eq. (\ref{deleqn}).
The results are shown for
two different choices of delay $k$. The black points indicate $\sigma < 10^{-15}$
and white indicates lack of synchronization.}
\label{linsta}
\end{figure}

\subsubsection{Effect of delays, $\beta > 0$}
Next, we consider the case $\beta > 0$. In this case, it
is not straightforward to analytically solve Eq. (\ref{deleqn}) to obtain
local stability criteria. We iterate Eq. (\ref{deleqn}) numerically
to estimate the value of $q$ [Eq. (\ref{locstab})] as a function of
$\alpha$ and $\beta$ for $\epsilon=0.1$.
In Fig \ref{linsta}, we present the numerical results and 
if $q<0$, we denote it by a black point (synchronization)
and for $q>0$ we denote it by white point (no synchronization).
The interesting feature is that the introduction of delays enhances
synchronization to some extent. For instance, at $\beta=0$, synchrony
requires $\alpha > 0.6$
but in the presence of delays approximately $\alpha > 0.4$
is sufficient for synchronization in the range $\sim 0.15 < \beta < 0.5$.
This is reminiscent of the recent results that indicate enhanced synchrony
due to presence of delays in coupled systems \cite{masoller,atay}.
As the strength of mutual coupling $\beta$ increases,
it will require even stronger
drive by $z$-CML to achieve synchronization. This can be qualitatively
seen in Fig \ref{linsta} where for $\beta > 0.5$, the minimum value of $\alpha$
required for synchronization increases with increase in $\beta$.
Notice again that at $\alpha=0$  there is no synchronization, even in the
presence of delays.

\begin{figure}[t]
\includegraphics*[width=8cm]{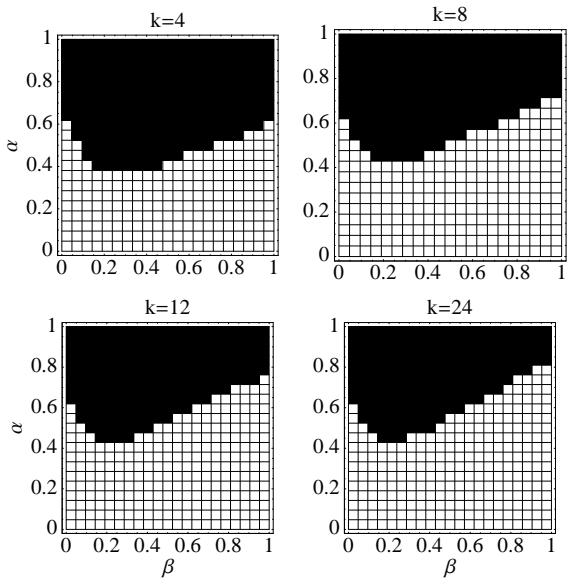}
\caption{Degree of synchronization $\sigma$, as a function of $\alpha$ and $\beta$,
obtained by numerically simulating CMLs in Eq. (\ref{cml-z}-\ref{cml-y}).
The results shown for four different choices of delay $k$.
The black points indicate $\sigma < 10^{-15}$
and white indicates lack of synchronization.}
\label{dosync}
\end{figure}

To confirm the linear stability analysis for $\beta > 0$, in particular the
results displayed in Fig \ref{linsta}, we simulate the CMLs
in Eqns. (\ref{cml-z}-\ref{cml-y}) as a function of $\alpha$, $\beta$ and $k$
and we display the degree of synchronization $\sigma$ in Fig \ref{dosync}.
The black points in the figure denote $\sigma < 10^{-15}$ and white points
correspond to lack of synchronization. This is shown for four different
choices of delays $k$. The CML simulations broadly agree with
the numerical estimates of $q$ shown in Fig. \ref{linsta} based on Eq. (\ref{deleqn}).
The features such as the enhancement in synchronization in the presence of
delays and optimal $\beta$ is clearly seen in these simulations. While a physical
explanation for enhanced synchrony and the optimal value of $\beta$ is not yet
clear, one plausible reason could be as follows;
For $\alpha=\beta=0$, we have two independent chaotic CMLs.
But, in the absence of only the driver CML, {\it i.e} with $\alpha=0$, the dynamics of the
CML system, $x_n(i)$ and $y_n(i)$, for $0.03 < \beta < 0.14$ settles mostly to a
periodic solution and for $\beta > 0.14$ it becomes increasingly chaotic. Thus, in
parameteric space, there
is a window of non-chaotic region flanked on either sides by predominantly chaotic dynamics.
The strength of the driver $\alpha$ required to synchronize the non-chaotic
dynamics is less than the one needed for chaotic solution.
This accounts for the dip around $\beta=0.1-0.2$ seen in Fig. \ref{linsta} and \ref{dosync}.

Further more, it is only to be expected that as $k$ increases synchrony would be
difficult to achieve and hence strong driving by $z$-CML will be needed to enforce
synchronization. Thus, if $\beta$ is held constant, the minimal $\alpha$
required to bring about synchronization increases as mutual coupling delay $k$
increases. At $k=\infty$, the delay is infinite and the $x$- and $y$-CMLs do not
communicate with each other on finite time scales. This scenario corresponds
to setting $\beta=0$, the absence of mutual coupling. Indeed if $k$ is larger
than the simulation times, we obtain similar results as shown in Fig. \ref{b0}.
Even though we use values of $k$ in multiples of 4, we emphasise that
the qualitative results remain unaltered for all even values of $k$. However,
for odd values of $k$, the synchronization region in $(\alpha,\beta)$-space
is smaller compared to those displayed in Fig \ref{dosync}.
A better analytical handle
on solutions of Eq. (\ref{deleqn}) will help understand the role of odd $k$.

\section{Robustness}
  How robust is this zero delay synchronization against parameter mismatches ?
This question is of practical importance since in real-life systems, be it the
EEG signals in the brain or the electronic circuits for encryption in
communications, most often the parameters remain mismatched. For the purposes
of this section, we will explicitly show the parameter dependence in the
CMLs; for instance, $x$-CML in Eq. (\ref{cml-x}) will be denoted by $x_{n+1}^i(a,\epsilon)$
and the local map will be denoted by, $f[x_n^i;a]$. We will consider the
quantity, to be called synchronization error,
\begin{equation}
S_{n+1}^i(a_1,\epsilon_1; a_2,\epsilon_2) = x_{n+1}^i(a_1,\epsilon_1) - y_{n+1}^i(a_2,\epsilon_2).
\label{sync-err}
\end{equation}
The synchronization time $T_{sync}$ is defined such that
$S_{n+1}^i(a_1,\epsilon_1; a_2,\epsilon_2)=0$ for all $n > T_{sync}$.
For most practical purposes, $T_{sync}$ should be
typically much smaller than the experimental times of interest.
We will consider the synchronization error in Eq. (\ref{sync-err}) and analytically
estimate the bounds on $S_{n+1}^i$ due to parameter
mismatches. In the case of identical synchronization without delay,
we have shown above that $S_{n+1}^i=0$ for all $i$ and for all $n > T_{sync}$.

Firstly, we note that identical synchronization persists,
i.e, $S_{n+1}(i)=0$ for reasonably large mismatches in
local map parameter $a$ between the driver CML and the driven CMLs.
In particular, the numerical simulations indicate that
synchronization is mostly independent of the coupling constant $\epsilon$
in $z$-CML. Hence the important effects arise due to mismatch in parameters
of the $x$- and $y$-CMLs, which we study below.
In the numerical simulations shown in this section, we have maintained
$\alpha=\beta=1$ and $k=26$.

\subsubsection{Mismatch in local map parameter}

We consider the effect of mismatch in the parameters of $x$ and $y$ CMLs.
First, we consider the case when parameters of $x$- and $z$- CML are
identical but there is a mismatch $\Delta a$ in local map
parameter between $x$- and $y$- CML.
Starting from Eqns. (\ref{cml-x},\ref{cml-y}), after some algebra,
we obtain $S_{n+1}^i$ without any approximation as,
\begin{equation}
S_{n+1}^i(a,\epsilon ; a-\Delta a, \epsilon) = S_{n+1}^i(a,\epsilon ; a, \epsilon) + \phi_1(\Delta a, \epsilon),
\label{mm1}
\end{equation}
where we have,
\begin{eqnarray}
\phi_1(\Delta a, \epsilon) &=& \Gamma \left( f[y_{n}^i;\Delta a] - \beta f[y_{n-k}^i;\Delta a] \right) + \nonumber \\
& & \frac{\epsilon}{2} \left( f[y_{n}^{i-1};\Delta a] + f[y_{n}^{i+1};\Delta a] \right).
\label{adep1}
\end{eqnarray}
For $n > T_{sync}$, we will have $S_{n+1}^i(a,\epsilon ; a, \epsilon)=0$ which is
the condition for zero lag synchronization and it has been numerically demonstrated
in the previous section. Notice that for logistic map $0 \le f[x_n;a] = a ~g(x_{n-1}) \le 1$,
where $g(x) = x(1-x)$. Hence the first term in Eq. (\ref{adep1})
is smaller compared to the second term and we have the approximate result that,
\begin{eqnarray}
S_{n+1}^i(a,\epsilon ; a-\Delta a, \epsilon) &\approx& \frac{\epsilon ~\Delta a}{2}
  \left( g[y_{n}^{i-1}] + g[y_{n}^{i+1}] \right) \nonumber \\
 & \le & \frac{\epsilon ~\Delta a}{2} g_{max} \approx \frac{\epsilon ~\Delta a}{2},
\label{adep2}
\end{eqnarray}
where $g_{max} = \mbox{max} \left( g[y_{n}^{i-1}] + g[y_{n}^{i+1}] \right) \sim O(1)$.
Thus, in case of mismatch in parameter $a$, the upper bound for synchronization
error is of $O(\epsilon ~\Delta a/2)$. This estimate can be compared with the
average synchronization error after evolving it for a sufficiently
long time as shown in Fig. \ref{fig6}(a,c). The root mean square deviation or the
standard deviation $\sigma_S(\Delta a)$ of $S_{n+1}^i(a,\epsilon ; a-\Delta a, \epsilon)$
is a suitable measure and is
obtained by numerically simulating Eqns. (\ref{cml-z}-\ref{cml-y}).
In Fig. \ref{fig6}(b), we show $\langle\sigma_S(\Delta a)\rangle$, the averaged
standard deviation over all lattice sites and we observe a good agreement with the
analytical result.

\subsubsection{Mismatch in coupling constant}

\begin{figure}[t]
\includegraphics*[width=8cm]{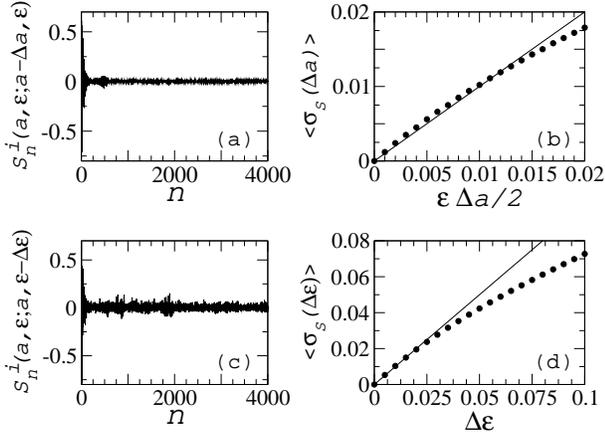}
\caption{(a) $S_n^i$ for $i=525$th lattice with $a=4.0,\epsilon=0.1$ and
$\Delta a=0.1$. (b) $\langle\sigma_S(\Delta a)\rangle$, the standard deviation of $S_n^i$
averaged over all lattice sites, as a function of $\Delta a$.
(c) $S_n^i$ for $i=525$th lattice with $a=4.0,\epsilon=0.1$ and
$\Delta \epsilon =0.05$ (d) $\langle\sigma_S(\Delta \epsilon)\rangle$
as a function of $\Delta \epsilon$.
Note that the synchronization error due to
parameter mismatch is bounded by estimates in Eqns. (\ref{adep2},\ref{epsdep1})
shown as solid lines in (b) and (d).}
\label{fig6}
\end{figure}

We consider the case when all the three CMLs in Eqns. (\ref{cml-z}-\ref{cml-y})
have the same chaos parameter $a$ but the local coupling strength in
$x$- and $z$-CML is $\epsilon$ and for $y$-CML it is $\epsilon + \Delta \epsilon$.
Typically, $\Delta \epsilon \ll 1$. Once again, we start from Eqns. (\ref{cml-x})
and (\ref{cml-y}) and we obtain,
\begin{equation}
S_{n+1}^i(a,\epsilon ; a, \epsilon - \Delta\epsilon) = S_{n+1}^i(a,\epsilon ; a, \epsilon) + \Delta\epsilon ~~\phi_2
\label{epsdep}
\end{equation}
\begin{eqnarray}
\phi_2 & = & -\frac{1}{3} \left( f[y_{n}^{i};a] + \beta f[x_{n-k}^i;a] + \alpha f[z_{n}^i;a]\right)  \nonumber \\
& & + \frac{1}{2} \left( f[y_{n}^{i-1};a] + f[y_{n}^{i+1};a] \right)
\end{eqnarray}
As before, $S_{n+1}^i(a,\epsilon ; a, \epsilon) = 0$ in Eq. (\ref{epsdep}) which
defines the synchronization state.
For the logistic map $0 \le f(x;a) \le 1$ and hence we have, $|\phi_2| \le 1$.
Thus, if the coupling parameters are mismatched the synchronization is still
present though it suffers an error whose bound is estimated to be,
\begin{equation}
S_{n+1}^i(a,\epsilon ; a, \epsilon - \Delta\epsilon) \le \Delta \epsilon.
\label{epsdep1}
\end{equation}
Fig. \ref{fig6}(d) shows that the numerically simulated synchronization error, quantified by
the average standard deviation $\langle\sigma_S(\Delta\epsilon)\rangle$ of 
$S_{n+1}^i(a,\epsilon ; a, \epsilon - \Delta\epsilon)$ over all lattice sites,
is always lesser than the analytical bound which is linear in $\Delta\epsilon$.

It must be remarked that in both the cases of parameter mismatches 
studied above synchronization
suffers an error that can be minimized by tuning $\Delta a$ or $\Delta \epsilon$.
In other words, for large mismatches ($\Delta a, \Delta \epsilon \gg 1$)
synchronization is completely lost as
seen from the trends in numerical results in Fig. \ref{fig6}(b,d).
Obviously, exactly identical synchronization without lag is recovered
if the mismatches $\Delta a$ and $\Delta \epsilon$ are zero. It is possible
that there can be mismatches in both $a$ and $\epsilon$. Proceeding as above,
we can obtain an estimate for the error bounds as,
\begin{eqnarray}
S_{n+1}^i(a,\epsilon ; a-\Delta a, \epsilon-\Delta \epsilon) & = & S_{n+1}^i(a,\epsilon ; a, \epsilon) +
\chi_1 ~\Delta a \nonumber + \\
& & \chi_2 ~\Delta\epsilon + \chi_3 ~\Delta a ~\Delta\epsilon,
\end{eqnarray}
where $\chi_1, \chi_2, \chi_3 \le 1$. Once synchronization is reached, 
$S_{n+1}^i(a,\epsilon ; a, \epsilon)=0$.
Thus, depending on the relative magnitude of $\Delta a,\Delta \epsilon$ and
$\Delta a ~\Delta \epsilon$, the dominant synchronization
error bound has linear dependence on one of these factors. 
In fact, the error analysis done above would hold good for any local map
of the form $f[x;a] = a~g(x)$. Experiments in
neuronal studies have reported examples of synchronization with error in
spatially distributed neurons \cite{konig}.
From the form of analytical estimates in Eqns. (\ref{mm1},\ref{adep2}) and
(\ref{epsdep1}), it might appear as though
$z$-CML has no role to play in synchronizing the coupled dynamics. The
contribution of $z$-CML enters the $x$- and $y$-CML through the mutual
coupling terms with the delay $k$.

\section{Conclusions}

We have shown through numerical simulations that two coupled map lattices,
say, $x$-CML and $y$-CML, which are coupled to one another with a delay
can display {\em isochronal} synchronization if they are driven by a
third CML. We have used periodic boundary conditions, {\it i.e.}, 
a ring type lattice for the CMLs in the results
presented here. The central result remains unaltered even if we use a different
boundary condition, e.g, the one way coupled map lattice with ring
or open type boundary conditions. While the isochronal synchronization
is achieved irrespective of the boundary conditions, the parametric
regimes in which this synchrony occurs depends on the boundary conditions.
Our results also indicate that there is a critical value for the
coupling strength $\epsilon$ above which the synchronization occurs. We have
analytically studied how the strength parameters $\alpha$ and $\beta$
and the delay $k$ affect isochronal synchronization. An interesting
feature is that the presence of delays leads to synchronization
in larger parametric regime when compared with the case with absence of
mutual coupling.

The original motivation for this work was to look for possible mechanisms that
could explain isochronal synchronization occuring in neuronal systems  \cite{eeg-rev}.
For simplicity, we considered one dimensional coupled map lattices with
nearest neighbour coupling even though they
are not known to be models for a collection of neurons in the brain. However,
we expect qualitatively similar results for models of neurons too \cite{ira}.
In general, higher
dimensional extensions of CMLs are possible and they display much richer
variety of collective properties like phase synchronized states and cluster
synchronization \cite{2dcml} which could modifiy the scenario presented in this work.
It would be interesting to study isochronal synchronization in higher dimensional
coupled map systems.

It is known that
two mutually coupled oscillators synchronize with one another. However, the
question of isochronal synchronization in mutually delay-coupled oscillators is
currently being actively pursued in view of its applications in
biological systems. The results discussed here provide one possible mechanism for
isochronal synchronization in delay-coupled spatially extended systems.
We have also obtained analytical estimates for bounds on errors due to
mismatches in parameters between the two CMLs and verified them in simulations.
We have also simulated this scheme with one way coupled map lattice \cite{will}
and qualitatively the results are the same as discussed above.
It would be interesting to study isochronal synchronization
in delay coupled physical systems and in their realistic models.
Further, other coupling topologies can also be implemented to study if
synchronized solutions such as the one discussed here is supported in them.
The results discussed in this work will help understand the effects
of delay coupling in spatially separated, extended physical systems.

\begin{acknowledgments}
One of us (SA) thanks Physical Research Laboratory for the internship during
which time part of this work was begun. We also thank the anonymous referee
for critical comments that helped improve the manuscript.
\end{acknowledgments}

\end{document}